\documentclass[10pt,twocolumn,letterpaper]{article}

\usepackage[pagenumbers]{cvpr} 

\usepackage[OT2,T1]{fontenc}

\definecolor{cvprblue}{rgb}{0.21,0.49,0.74}
\usepackage[pagebackref,breaklinks,colorlinks,allcolors=cvprblue]{hyperref}

\usepackage{soul}
\usepackage{float}
\usepackage{graphicx}
\usepackage{amsmath}
\usepackage{amssymb}
\usepackage{afterpage}
\usepackage{scalerel,xparse}
\usepackage{capt-of,etoolbox}
\usepackage{booktabs}
\usepackage{mathtools}
\usepackage{array}
\usepackage{soul}
\newcolumntype{P}[1]{>{\centering\arraybackslash}p{#1}}
\usepackage{float}
\usepackage{multirow}
\usepackage{float}      
\usepackage{afterpage}
\usepackage{scalerel,xparse}
\usepackage{capt-of,etoolbox}
\usepackage{tikz}
\usepackage{algorithm}
\usepackage{algpseudocode}
\usepackage[pagebackref,breaklinks,colorlinks]{hyperref}
\usepackage{subcaption}

\newcolumntype{N}{>{\centering\arraybackslash\footnotesize}m{.5in}}
\newcolumntype{G}{>{\centering\arraybackslash}m{34pt}}
\usepackage[capitalize]{cleveref}

\usepackage{listings}
\usepackage{xcolor}

\newcommand{\myparagraph}[1]{\noindent\textbf{#1}}

\lstdefinelanguage{Python}{
  keywords={def, return, if, else, for, while, in, import, from, as, try, except, with, lambda, True, False, None},
  keywordstyle=\color{blue},
  identifierstyle=\color{black},
  sensitive=true,
  comment=[l]\#,
  commentstyle=\color{gray},
  stringstyle=\color{green},
  morestring=[b]',
  morestring=[b]",
}

\lstdefinelanguage{JavaScript}{
  keywords={const, let, var, function, return, if, else, for, while, in, try, catch, await, async, new, this, require},
  keywordstyle=\color{blue},
  identifierstyle=\color{black},
  sensitive=true,
  comment=[l]//,
  commentstyle=\color{gray},
  stringstyle=\color{green},
  morestring=[b]',
  morestring=[b]",
}

\def\paperID{*****} 
\def\confName{CVPR}
\def\confYear{2025}

\title{Accelerating Blockchain Scalability: New Models for Parallel Transaction Execution in the EVM}

\author{Souradeep Das\\
University of California, Berkeley\\
Berkeley\\
{\tt\small souradeep@berkeley.edu}
\and
Jonas Bäumer\\
University of California, Berkeley\\
Berkeley\\
{\tt\small jonasbaeumer@berkeley.edu}
\and
Konpat Preechakul \\
University of California, Berkeley\\
Berkeley\\
{\tt\small konpat@berkeley.edu}
\and
Riddhi Patel\\
University of California, Berkeley\\
Berkeley\\
{\tt\small riddhipatel@berkeley.edu}
\and
Jefferson Jinchuan Li\\
University of California, Berkeley\\
Berkeley\\
{\tt\small jinchuanli@berkeley.edu}
}

\begin{document}
\maketitle

\begin{abstract}
As the number of decentralized applications (dApps) and users on Ethereum grows, the ability of the blockchain to efficiently handle a growing number of transactions becomes increasingly strained. Ethereum’s current execution model relies heavily on sequential processing, meaning that operations are processed one after the other, which creates significant bottlenecks to future scalability demands. While scalability solutions for Ethereum exist, they inherit the limitations of the EVM, restricting the extent to which they can scale.
This paper proposes a novel solution to enable maximally parallelizable executions within Ethereum, built out of three self-sufficient approaches. These approaches include strategies in which Ethereum transaction state accesses could be strategically and efficiently predetermined, and further propose how the incorporation of gas-based incentivization mechanisms could enforce a maximally parallelizable network.
\end{abstract}

\section{Introduction}
Imagine a world-class highway system where only one car is allowed to drive at a time. This is Ethereum today—a revolutionary platform constrained by outdated traffic rules. Despite its transformative impact as a decentralized blockchain powering thousands of applications, from financial services to digital collectibles, Ethereum faces critical limitations. At its core lies the Ethereum Virtual Machine (EVM), which processes transactions sequentially to ensure security and consistency \cite{Buterin2014-uw}. While this approach has been foundational in establishing trust and reliability, it imposes significant scalability challenges, resulting in high fees and network congestion as demand grows.

These scalability challenges have spurred the development of numerous Layer 2 (L2) solutions, such as Optimistic Rollups, ZK-Rollups, and Plasma \cite{Thibault2022-lp, Poon2017-ub}. By offloading or batching transactions off-chain, these solutions alleviate some of Ethereum's bottlenecks, enabling higher throughput and lower costs. However, they fail to address the root cause: Ethereum's base layer's inability to process transactions in parallel. This limitation constrains the network's efficiency, inflates transaction costs, and restricts developers from fully realizing blockchain's potential for high-throughput applications.

The problem is further exacerbated in EVM-based layer 2 solutions, where a small group of specialized nodes--known as sequencers--are responsible for executing transactions. The throughput of these layers remains tightly bound to the sequential nature of the EVM's execution. For example, during periods of high demand, transaction fees can skyrocket. On May 1, 2022, the average daily gas price rose to \$196.638, making Ethereum unaffordable to many users \cite{UnknownUnknown-tw}. This underscores a critical inefficiency in the EVM: its inability to distinguish between parallelizable transactions and those that require sequential execution.

To address this, we propose a comprehensive solution built on three self-sufficient pillars, designed to enable scalable and parallel execution within Ethereum. First, we restructure Ethereum's gas mechanics to factor in the opportunity cost of non-parallelizable transactions, creating incentives for efficient block construction. Second, we introduce developer-implemented access boundaries that define the dependencies of each transaction. Third, we propose a method for generating and incorporating (user-provided) access lists to facilitate safe and efficient parallel execution. While the second and third pillars enable parallelization directly, true scalability can only be achieved when restructured gas incentives nudge users and block builders toward more optimal transaction groupings.

By implementing these strategies in unison, our proposal aims to transform Ethereum into a more scalable platform capable of meeting the growing demands of decentralized finance (DeFi) and beyond. In doing so, we not only address the current inefficiencies but also lay the groundwork for Ethereum's evolution into a truly high-throughput, parallelized blockchain.

\section{Related Works}
The approaches proposed in this paper are novel and no similar approach or construction to parallelization exists. However, our research benefits from several fundamental parallelization approaches proposed before. 

\myparagraph{EIP 2930 \cite{Swende2020-qv}} is implemented and active on the Ethereum mainnet right now. Although intended for stateless clients, this work is a great precursor to implementing access lists on Ethereum. While we necessarily need exhaustive access lists, the “optional access lists” TX option potentially can be repurposed to support our proposal's smart access lists.  

\myparagraph{EIP 648 \cite{ButerinUnknown-ij}.} Vitalik’s proposal of extending optional access lists to help with parallelization. However, as this approach relies on estimations based on the current state before transaction execution, there is no guarantee that the access traces will remain consistent during runtime. If the contract state changes between simulation and actual execution, the access list becomes invalid.

\myparagraph{Block STM \cite{Gelashvili2022-yq}.} One of the speculative concurrency methods for parallelization, where transactions are optimistically executed in parallel until there is a collision. If there is a collision, the transactions are re-queued to be executed sequentially. Block STM uses concepts of software transactional memory to create dynamic dependency graphs and a smart scheduler. However, speculative concurrency has proved to be less effective as the size of the chain grows \cite{Saraph2019-st}.

\myparagraph{EIP-7782 \cite{Adams2024-rt}} is a proposal to reduce Ethereum's slot time from 12 seconds to 8 seconds, aiming to increase transaction throughput by approximately 33\%. Our project is a perfect complement for it, adding a layer of reliability to the fact that slot times could be lower, and parallel execution will assist in that.
There are 3 phases in terms of time allocated to each slot (build, propose, aggregate sigs) (the 3rd step includes multiple network propagation messages). We hypothesize that we can shorten block-building time since execution time goes down. Also, validation time goes down given validators can execute faster. With that, even with a constant propagation delay, we think we can still reduce the slot time as an effect of parallelization. 

\myparagraph{Predict-al \cite{AshuUnknown-qy}} is a static bytecode analysis tool that can be used to predict access lists. While we do not propose a tool, tools similar to this could be of assistance to developers in generating method access boundaries for their smart contracts.

Specifically, most prior research and work on the parallelization of transactions in the EVM fall majorly into two categories: a) speculative concurrency, and b) access-list-based execution. Speculative concurrency can range from being simple optimistic execution to being enforced through the separation of nodes. However, these approaches tend to show less benefits over a sequential model in a practical setting. Access list approaches on the other hand work with prefetched/provided data for determining the most effective execution. However, access lists are hard to accurately determine before the transaction is actually executed, and enforcing a strict access list requirement is still hard from a user's or a commonly used wallet’s perspective.

Some alternative perspectives to parallelize beyond these strategies have also been proposed in the past, specifically, Garamvölgyi et al.’s approaches to affect specific cases like counters and commutative instructions \cite{Garamvolgyi2022-ax}. Given the nature of counters as a storage medium, their approaches included sharding these states and creating a new (deterministic) scheduler, and these strategies could complement our approaches.

\section{Approach}

To enhance the parallelization of transactions in the EVM, and augment scalability limits we focus on a \textbf{threefold approach}. Each component is designed to work synergistically and complementarily within our solution, aiming to maximize Ethereum's (and the EVM’s) parallelization potential while also providing strong incentives for adoption. Our solution will include the following:

\subsection{Gas Incentivization Mechanics}

We aim to restructure gas mechanics in a way that would a) incentivize users to strive for forming more parallelizable transactions and b) enforce block proposers (builders) to form a more efficient organization of blocks. In our re-imagined approach:
\begin{enumerate}
    \item Transaction fee mechanics are updated (in compliance with EIP-1559 \cite{ButerinUnknown-xq}), such that a portion of the base fee goes to the validators if a transaction is parallelizable, while the other portion is burned as per current practice. 
    \item Every transaction will be assumed to be parallelizable within a certain time, and a transaction will wait in the mempool with the hope of a proposer to pick it up, to be able to execute in parallel. 
    \item Each block is split into sections with identifiers that indicate transactions that can be executed in parallel and ones that need to be executed in sequential order. Proposers that can include transactions in the parallel section of the block end up getting a portion of the base fee for those transactions, incentivizing proposers to form more efficient blocks. (This also changes the narrative of Ethereum to being pro-validators (miners) where the validator is doing effective work by making more parallelizable blocks and gets rewards based on their performance).
    \item The incentives for the users (to provide a strict access list, mentioned below) or devs to create more efficient contracts is to have transactions included faster at a substantially low general cost if it is parallelizable.
    \item Transaction inclusion is promised but the time of inclusion depends on incentives. If a certain number of blocks has passed, while the proposer hasn’t still included the transaction (because proposers do not get a portion of the base fee if the transaction cannot be put on the parallel section of the block) - then a proposer is bound to include it sequentially.
    \item If a user wants to include their transaction faster (while it is not parallelizable, or hard to parallelize) they can do so by paying extra-priority fees on their transaction. Thus, users provide/make up for the incentive part of the transaction in this case.
    \item While the priority fees for a certain transaction cannot be fully predicted, a reasonable range of priority fees for a certain transaction of a certain smart contract can be derived from historical data. Such information can be provided by a third-party service coupled with a user-friendly interface (on the wallets) (note- that transaction inclusion is still guaranteed and transaction can be resubmitted with higher fees anytime).
\end{enumerate}

Overall, to restate, this approach brings about the following changes:
\begin{enumerate}[label=(\alph*)]
    \item More parallelizable transactions are incentivized because they are likely to be picked faster by the proposer,
    \item If a user is willing to pay extra priority fees then they have a higher chance to get faster inclusion (even sequentially) - since it makes up for the proposer's profit.
\end{enumerate}

\myparagraph{Incentives for proposers to maximize fees.}
While giving away a portion of the base fee (instead of burning) has been envisioned to be the incentive for proposers including parallel transactions, this incentive can take different forms and be generated in more than one way. 
Parallelizable transactions mean faster total block process time, even though a large part of which includes time for propagation and validator signature aggregation, because the execution portion of the block process is saved due to parallelization. 
This allows us to reduce the block time potentially from 12s to 8s (EIP 7782 \cite{Adams2024-rt}) where the proposers who capitalize better on the parallelizability of the transactions can fit more into a block and are rewarded with more fees per block.
This forms a natural way to incentivize the proposers and validators even without explicitly being rewarded a portion of the base fee.

\subsection{Method Access Boundaries}

A critical portion of our parallel execution approach is incentivizing proposers to form more efficient blocks. The second and third pillar of our solution proposes novel approaches in which the proposer (or even the EVM in a self-sufficient independent model) can identify the most efficient parallelizable set of transactions. 

To enable parallel execution of transactions in Ethereum, we propose introducing  Method Access Boundary as a sidecar structure for smart contracts. This structure will need to be specified/provided by the dev/maintainer of the specific smart contract and it is a database that defines the boundaries of internal/external state and storage access for each method of the contract

Each transaction will then be assigned the default boundaries based on the contract method they intend to invoke, and in that way, each transaction can be represented as a node in a graph, where edges indicate dependencies or conflicts between transactions. If two transactions are connected by an edge, they access a shared state/storage and thus cannot be executed in parallel. The challenge for the proposer then becomes finding the largest independent subset of transactions — those that don’t share dependencies and can be executed simultaneously.

To further elaborate on the details of the approach:
\begin{enumerate}
     \item Using the Method Access Boundary, we can pre-define a "scope" for each contract method, specifying the exact state/storage elements it accesses. Transactions within non-overlapping scopes are independent and can be executed in parallel.
     \item With this structure in place, the gas pricing model can reward users for submitting transactions with minimal dependencies(indirectly depending on the contract), thereby promoting independent, parallelizable transactions.
     \item For transactions directed to methods that do share dependencies, priority fees can be adjusted dynamically based on the degree of dependency (or prediction given the boundary is a statically predefined list, creating a market-based approach to balance parallel and sequential transactions.
\end{enumerate}

Future Iterations of Method Access Boundaries could introduce dynamic elements, adapting boundaries in real-time to reflect changes in contract states. Complex cross-contract dependencies are another area of focus, with composite boundaries potentially enabling parallelization for transactions spanning multiple contracts.

For developer enablement, to generate these method access boundaries, tools, and frameworks are critical. And, it is expected that the smart contract developer to provide the boundaries post-deployment. This can be assumed to be a similar step to contract verification in terms of complexity. Pre-deployment tools could be created to assist developers or users to identify and generate boundaries for each method of the smart contract. 

\subsection{Smart Access Lists}

Smart Strict Access Lists introduce an advanced method to enforce precise and efficient access list creation, dynamically generated for each transaction and specified within the transaction object, ensuring availability during execution. By requiring strict access lists with intelligent creation methods, this approach enhances the EVM’s ability to determine which transactions can be executed in parallel, thereby improving efficiency. This method also leverages ERC 4337 Smart Accounts to enable the collaborative creation of access lists for transactions.

The objective is to assist clients in generating accurate “access lists” through a Smart Entity, an automated mechanism that understands contract interactions in depth. Generating or expecting state interactions by simulating the transaction (similar to \texttt{eth\_estimateGas}) is possible. However, creating access lists that remain valid at runtime requires a substantial understanding of the smart contract, which is where the Smart Entity plays a crucial role.

\myparagraph{Access List Suppliers.} The Smart Entity, as an access list supplier, operates by:
\begin{enumerate}
    \item Coordinating access list creation and validation through signatures.
    \item Employing rule-based heuristics, drawn from developer knowledge, to create access lists that remain stable at runtime.
\end{enumerate}

\myparagraph{Process Flow}
\begin{enumerate}
    \item User Action: A user initiates a transaction and signs it.
    \item Entity Validation: The Smart Entity receives the transaction, appends the necessary access list, signs it again (double signature), and coordinates its execution.
    \item Heuristic Use: The entity uses rule-based heuristics from developers to anticipate access requirements and minimize the chance of invalid lists.
\end{enumerate}

\myparagraph{Self Sufficiency of Approaches.} While we propose a solution that involves three pillars (approaches) to provide a cumulative system that values and maximizes parallelization in Ethereum, each of these approaches are self-sufficient in enforcing parallelization. For instance, a chain that only implements method access boundaries or smart access lists could still parallelize the existing inflow of transactions by having the client (EVM indirectly) utilize this as a part of processing transactions. Combining these three approaches in our proposed solution provides a maximal parallelization opportunity. In other words, it takes the solution from parallelizing the existing inflows to incentivizing parallelizable block creation.

\myparagraph{Sandboxed Execution.} In our proposed solution, the proposer identifies and groups transactions that can be executed parallelly for the maximum profit. In order to do this - it relies on either method access boundaries or smart access lists. Smart access lists have a preference over method access boundaries, hence, if smart access lists are specified they are selected over method access boundaries for determining parallelizable sets of transactions. 

There is a chance, however, that smart access lists provided with a transaction are incomplete or invalid, in which case the proposer will end up having an invalid state - and this will only be caught during execution. To protect other transactions in the same block (set) with valid sets (and not have to revert them because of the existence of a transaction with an invalid access list), we propose sandboxed execution - where if smart access lists are specified - the transaction will be limited to utilizing only the specified state/storage values while executing the transaction. Any access outside the specified list in the course of execution will lead to that specific transaction failing.

\myparagraph{Forward-looking Access Lists.} Dynamic adjustments and validations are integral to the evolution of Smart Access Lists. The Smart Entity could implement a feedback loop that analyzes runtime execution data to refine its heuristics continuously. This would allow access lists to adapt dynamically to contracts with evolving states or unforeseen interactions, minimizing execution failures.

As Ethereum continues to evolve, Smart Access Lists could integrate seamlessly with proposed upgrades or Layer 2 solutions, ensuring compatibility and scalability since they do not require network level changes.

\section{Evaluation}
We aim to demonstrate the potential for parallelizing Ethereum transactions using historical data. Under our proposals, the block proposers are tasked to select and schedule the transactions from those available in their mempools before including them into a block. We analyze the Ethereum transactions from blocks 21,259,000 to 21,259,100, with approximately 12k transactions. The code used for obtaining the data is provided in Supp.

Simulating this scenario from historical data is challenging mainly because we do not have the record of the actual mempool. We adopted to simulate the mempool from the information we have about the transactions in the blockchain, based on two requirements: while the mempool can be much larger than a block size, the mempool should only contain temporally local transactions. To maintain temporal locality, our mempools are composed of three times the average block size number of consecutive transactions.

The process of scheduling is to group transactions into parallelizable groups wherein each group all transactions are guaranteed to have non-overlapping dependencies. It is worth noting that the optimal scheduling problem can be reduced to the Maximum Independent Set (MIS) problem in graph theory, which is known to be NP-hard \cite{Lawler1980-oq}. Therefore, in this section, we will use a simple greedy scheduling algorithm (provided in Supp.) to serve as a lower bound.

\subsection{Question 1: To what degree transactions can be parallelized?}

\begin{figure}
\centering
\includegraphics[width=1\linewidth]{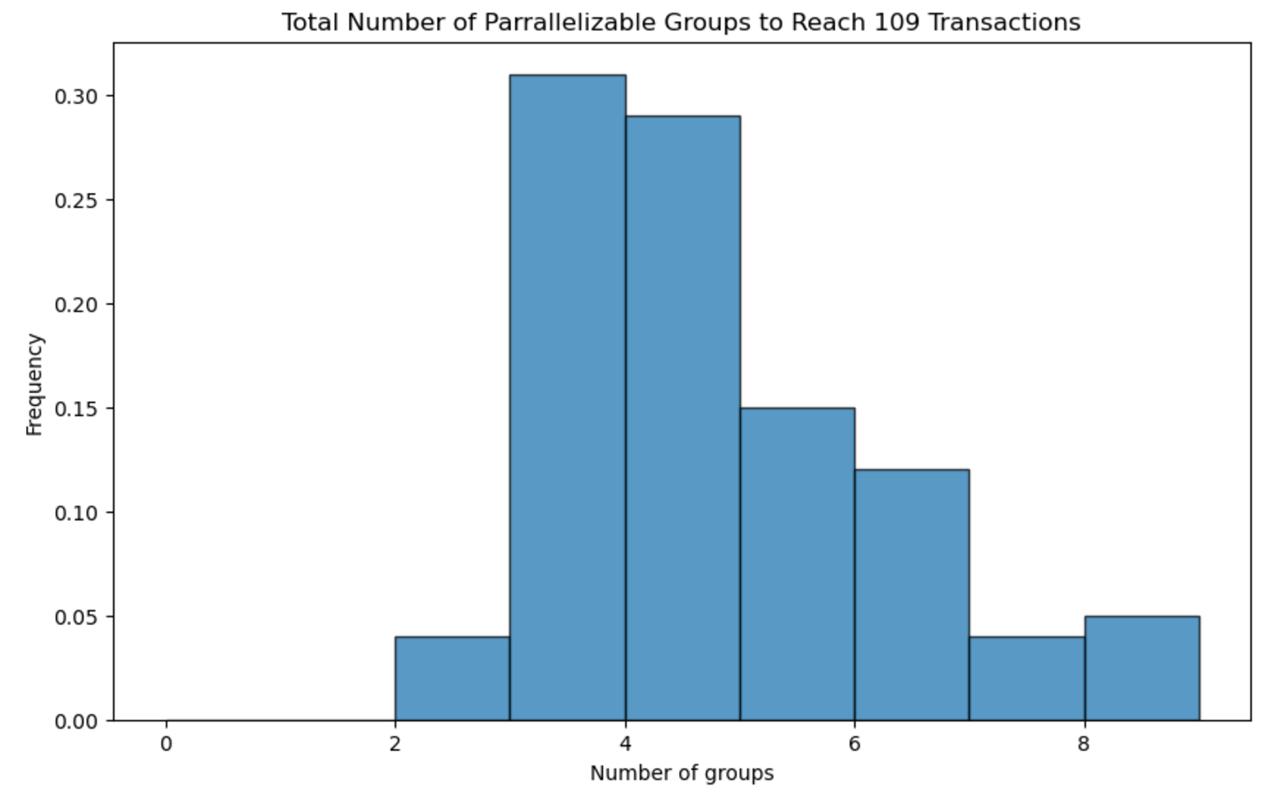}
\caption{On average, only 4 parallelizable groups are needed for 109 transactions within a block. This is a surprising result that demonstrates the significant potential of parallelization.}
\label{fig:group_size}
\vspace{-1em}
\end{figure}

Our first experiment simulates the creation of a block from the mempool. Consecutive transactions are randomly selected to simulate the mempool, and parallelizable groups are generated from it. Considering the constraints of block size, we sort the groups by the number of transactions they contain, in descending order, and fill the block with the groups until the average block size is reached. The number of groups required to form a block is then recorded. A lower number of groups indicates a higher degree of parallelization. This process is repeated 100 times, and the results are shown in Fig. \ref{fig:group_size}. 

A potential limitation of this experiment is its focus on optimizing a single block, which may not reflect long-term characteristics. As the most parallelizable transactions are prioritized in the first block, subsequent blocks may require more groups to execute the remaining transactions.

\subsection{Question 2: What is the sustainable degree of parallelization?}

\begin{figure}
\centering
\includegraphics[width=1\linewidth]{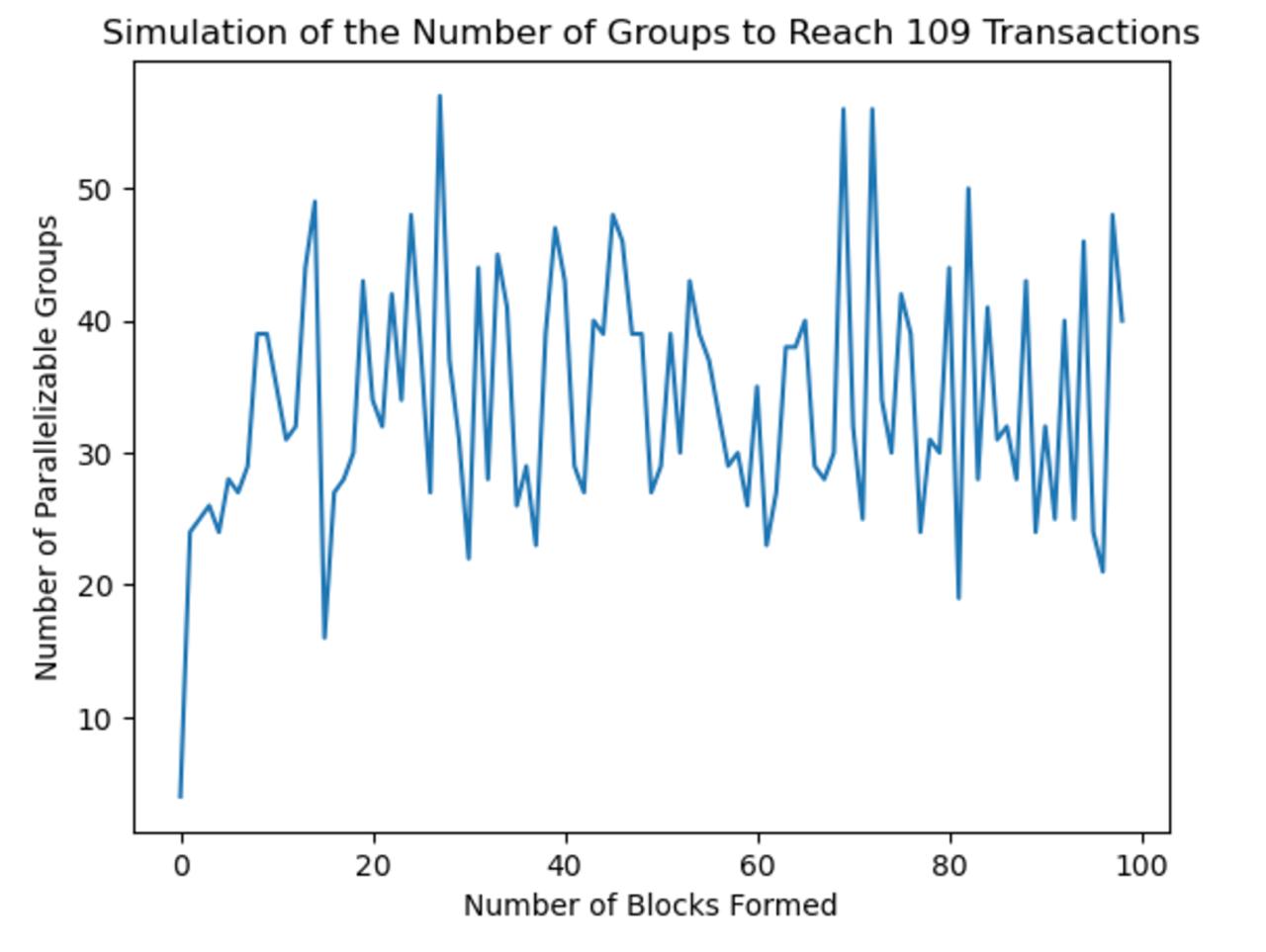}
\caption{Initially, the number of groups required is small, but it quickly increases and oscillates between 20 and 50 groups. The median number of groups required is 32 for 109 transactions, demonstrating that our approach achieves approximately three times the efficiency of traditional sequential execution in the long term.}
\label{fig:group_time}
\vspace{-1em}
\end{figure}

To address this concern, we conduct a second experiment to evaluate the effectiveness of parallelization over the long term. Using data from 100 consecutive blocks, we begin by forming a block from the mempool containing transactions from the first three blocks. Transactions from the fourth block are then added to the mempool, and a new block is formed. We repeat this process to simulate the formation of blocks in a realistic setting. The number of parallelizable groups required for each block is shown in Fig. \ref{fig:group_time} which suggests that as the most parallelizable transactions are taken to form blocks, the transactions left are less likely to be as parallelizable leading to a larger number of parallelizable groups. 

It is worth noting that we do not have access to the real mempool, which contains many more transactions. It is a possibility that there are enough activities in the real mempool such that a relatively small number of groups can always be formed.

\subsection{Question 3: How much does parallelization improve throughput?}

The goal of parallelization is to increase execution speed, translating into higher throughput for the Ethereum blockchain. This benefits users through reduced transaction fees and benefits proposers by enabling them to process more transactions per second. Proposers can afford to accept smaller fees per transaction while earning higher total fees due to the increased transaction volume.

Our analysis indicates that the median sustainable number of parallel groups per block (containing 109 transactions) is 32. With this configuration, the degree of parallelization is approximately $109 / 32 \approx 3.4$. To quantify the throughput improvement, we model the total block time as the sum of block propagation time ($p$) and block computation time ($c$): 
\begin{align}
    \text{Total block time} = p + c \, \text{seconds}.
\end{align}

The throughput is given by:
\begin{align}
    \text{Throughput} = \frac{\text{Block size}}{p + c}.
\end{align}

With computation reduced by a median factor of $3.4$ through parallelization, the new throughput becomes:
\begin{align}
    \text{New Throughput} = \frac{\text{Block size}}{p + \frac{c}{3.4}}.
\end{align}

Assuming a total block time of 12 seconds split evenly among $p = c = 6$ seconds, the new throughput improves from 9.08 to 14.04 transactions per second, representing a 1.54x increase.





\section{Conclusion}
We presented in this report a three-fold proposal to help synergistically enable parallelization in Ethereum. Our proposal incentivizes the users to annotate their transactions with access lists, and the proposers and validators to utilize that information to come up with a clever transaction scheduling algorithm to speed up execution through parallelization. Our proposal naturally creates a virtuous cycle where the users, and thereby developers, produce more parallelizable transactions to save fees which leads to a more efficient Ethereum network as a whole.

{
    \small
    \bibliographystyle{ieeenat_fullname}
    \bibliography{paperpile}
}


\clearpage
\maketitlesupplementary
\appendix

\section{Python Code for Overlap Pairs and Group Generation}

\subsection{Function to Generate Overlap Pairs}
\begin{lstlisting}[language=Python]
def generate_overlap_pairs(transaction_address_map):
    """
    Generate overlap pairs between transactions based on address overlaps.
    """
    overlap_pairs = {}  # Initialize an empty dictionary to store overlapping pairs
    transaction_nums = list(transaction_address_map.keys())  # Get a list of all transaction IDs

    # Compare each pair of transactions for overlaps
    for i in range(len(transaction_nums)):
        for j in range(i + 1, len(transaction_nums)):
            transaction_a, transaction_b = transaction_nums[i], transaction_nums[j]

            # Check if the two transactions have overlapping addresses
            if transaction_address_map[transaction_a] & transaction_address_map[transaction_b]:
                # Store the overlapping pair and their intersecting addresses
                overlap_pairs[(transaction_a, transaction_b)] = (
                    transaction_address_map[transaction_a] & transaction_address_map[transaction_b]
                )

    return overlap_pairs  # Return the dictionary of overlapping pairs
\end{lstlisting}

\subsection{Function to Generate Optimized Groups}
\begin{lstlisting}[language=Python]
def generate_groups(transaction_address_map, overlap_pairs):
    """
    Generate optimized groups of transactions where each group contains parallelizable transactions.
    """
    # Initialize degrees for each transaction (number of overlaps)
    transaction_degrees = {transaction: 0 for transaction in transaction_address_map}

    # Calculate degrees based on overlap pairs
    for transaction_a, transaction_b in overlap_pairs.keys():
        transaction_degrees[transaction_a] += 1
        transaction_degrees[transaction_b] += 1

    # Sort transactions by degree in descending order (higher overlap transactions first)
    sorted_transactions = sorted(transaction_degrees.keys(), key=lambda x: transaction_degrees[x], reverse=True)

    # Initialize the list of optimized groups
    optimized_groups = []

    # Assign each transaction to a group
    for transaction_num in sorted_transactions:
        added = False
        # Try to add the transaction to an existing group
        for group in optimized_groups:
            # Check if the transaction overlaps with any transaction in the current group
            if not any(transaction_address_map[transaction_num] & transaction_address_map[other_transaction] for other_transaction in group):
                group.add(transaction_num)  # Add to group if no overlap
                added = True
                break
        # If no suitable group is found, create a new group for the transaction
        if not added:
            optimized_groups.append({transaction_num})

    return optimized_groups  # Return the list of optimized groups
\end{lstlisting}

\section{JavaScript Code for Storage Access Analysis}
\begin{lstlisting}[language=JavaScript]
const { providers } = require('ethers');
require('dotenv').config();
const fs = require('fs');

const l2Provider = new providers.JsonRpcProvider('https://nd-422-757-666.p2pify.com/0a9d79d93fb2f4a4b1e04695da2b77a7/');

let noOfErrors = 0;

async function main() {
    console.log('Finding Storage accesses');

    let globalList = [];
    for (var i = 21259000; i <= 21259100; i++) {
        try {
            const blockNum = i;
            const block = await l2Provider.getBlock(blockNum);
            const ts = block.timestamp;
            const transactions = block.transactions;

            for (const txHash of transactions) {
                let addresses = [];
                try {
                    const trace = await l2Provider.send('debug_traceTransaction', [txHash, {
                        tracer: '{' +
                            'retVal: [],' +
                            'afterSload: false,' +
                            'callStack: [],' +

                            'byte2Hex: function(byte) {' +
                            '  if (byte < 0x10) ' +
                            '      return "0" + byte.toString(16); ' +
                            '  return byte.toString(16); ' +
                            '},' +

                            'array2Hex: function(arr) {' +
                            '  var retVal = ""; ' +
                            '  for (var i=0; i<arr.length; i++) ' +
                            '    retVal += this.byte2Hex(arr[i]); ' +
                            '  return retVal; ' +
                            '}, ' +

                            'getAddr: function(log) {' +
                            '  return this.array2Hex(log.contract.getAddress());' +
                            '}, ' +

                            'step: function(log,db) {' +
                            '   var opcode = log.op.toNumber();' +

                            '   if (opcode == 0x54) {' +
                            '     this.retVal.push("SLOAD " + ' +
                            '        this.getAddr(log) + ":" + ' +
                            '        log.stack.peek(0).toString(16));' +
                            '        this.afterSload = true; ' +
                            '   } ' +

                            '   if (opcode == 0x55) ' +
                            '     this.retVal.push("SSTORE " +' +
                            '        this.getAddr(log) + ":" + ' +
                            '        log.stack.peek(0).toString(16));' +

                            '},' +

                            'fault: function(log,db) {this.retVal.push("FAULT: " + JSON.stringify(log))},' +

                            'result: function(ctx,db) {return this.retVal}' +
                        '}'
                    }]);

                    for (let i = 0; i < trace.length; i++) {
                        let res = trace[i].match(/(\b[a-f0-9]{40}\b)/g);
                        addresses.push(res[0]);
                    }
                    let uniqueAddr = [...new Set(addresses)];

                    const addrObj = { "blockNum": blockNum, "ts": ts, "addresses": [] };
                    for (let j = 0; j < uniqueAddr.length; j++) {
                        addrObj.addresses.push({ "addr": uniqueAddr[j], "read": [], "write": [] });
                    }

                    for (let k = 0; k < addrObj.addresses.length; k++) {
                        for (let l = 0; l < trace.length; l++) {
                            if (addrObj.addresses[k].addr == (trace[l].match(/(\b[a-f0-9]{40}\b)/g))[0]) {
                                if ((trace[l].match(/SLOAD|SSTORE/g))[0] == 'SLOAD') {
                                    addrObj.addresses[k].read.push((trace[l].match(/[^:]*$/g))[0]);
                                } else if ((trace[l].match(/SLOAD|SSTORE/g))[0] == 'SSTORE') {
                                    addrObj.addresses[k].write.push((trace[l].match(/[^:]*$/g))[0]);
                                }
                            }
                        }
                    }

                    for (let k = 0; k < addrObj.addresses.length; k++) {
                        addrObj.addresses[k].read = [...new Set(addrObj.addresses[k].read)];
                        addrObj.addresses[k].write = [...new Set(addrObj.addresses[k].write)];
                    }

                    console.log(addrObj);

                    fs.appendFileSync('txDataFinal.json', JSON.stringify(addrObj));
                    globalList.push(addrObj);
                } catch (txErr) {
                    console.log(`Error processing transaction ${txHash}:`, txErr);
                    noOfErrors++;
                }
            }
        } catch (blockErr) {
            console.log(`Error processing block ${i}:`, blockErr);
            await new Promise(r => setTimeout(r, 2000));
        }
    }

    console.log('no of errors: ', noOfErrors);
    console.log('Global addr: ', globalList);
    fs.appendFileSync('txDataFinal.json', JSON.stringify(globalList));
}

main();
\end{lstlisting}

\end{document}